**Introducing physics-informed generative models for targeting structural novelty in the exploration of chemical space**

A. Vasylenko[1], F. Ottomano[2], C. M. Collins[3], R. Savani[2], M. S. Dyer[1,3], M. J. Rosseinsky[1,3]

[1] Department of Chemistry, University of Liverpool, L69 7ZD, UK
[2] Department of Computer Science, University of Liverpool, L69 3DR, UK
[3] Leverhulme Research Centre for Functional Materials Design, Materials Innovation Factory, University of Liverpool, L69 7ZD, UK

## Abstract

Discovering materials with new structural chemistry is key to achieving transformative functionality. Generative artificial intelligence offers a scalable route to propose candidate crystal structures. We introduce a reliable low-cost proxy for structural novelty as a conditioning property to steer generation towards novel yet physically plausible structures. We then develop a physics-informed diffusion model that embeds this descriptor of local environment diversity together with compactness as a stability metric to balance physical plausibility with structural novelty. Conditioning on these metrics improves generative performance across diffusion models, shifting generation away from structural motifs that dominate the training data. A chemically grounded validation protocol isolates those candidates that combine plausibility with structural novelty for physics-based calculation of energetic stability. Both the stability and the novelty of candidates emerging from this workflow can however change when the full potential energy surface at a candidate composition is evaluated with crystal structure prediction (CSP). This suggests a practical generative–CSP synergy for discovery-oriented exploration, where AI targets physically viable yet structurally distinct regions of chemical space for detailed physics-based assessment of novelty and stability.

## Main

Materials underpin and drive technology. The properties of materials are defined by their structures and compositions, so the experimental realisation (that is, the discovery) of new structural chemistry beyond the limits of existing databases is essential in the search for unprecedented functionalities, from energy storage to quantum-information platforms[1,2]. For example, the discovery of a new structure type in $Li_{10}GeP_2S_{12}$ led to solid Li-ion electrolytes with liquid-like conductivity[3]. The chemical space of crystalline solids is jointly defined by structure and composition, where new functionalities emerge in regions that remain experimentally unexplored. Compositions and structures are therefore deeply interdependent, and exploring this coupled space efficiently remains a major challenge for materials discovery. The aim is to identify those compositions that support experimentally accessible new structural chemistry, combining underexplored structural motifs with physical plausibility.

There has been recent progress in scalable methods for algorithmic materials generation[4–11]. Generative AI models provide new instances based on training data[12] and are typically optimised and benchmarked for distributional fidelity[13,14] – how closely their outputs reproduce the statistical patterns in the training set – rather than for extrapolative discovery. There is then an opportunity to produce structurally diverse and physically plausible candidates outside well-charted regions of chemical space[15,16], thus maximising impact within experimental discovery workflows[17]. A key challenge is therefore to move out-of-distribution in structural chemistry while retaining energetic plausibility. A generative model that combines chemistry-informed conditioning with targeted sampling of viable novel structural motifs could then guide the exploration of chemical space.

Crystal structure prediction (CSP) is a class of global optimisation methods[18–20] that explore the potential energy surface at a given composition to search for low-energy configurations. It is an established approach to propose new plausible candidate compounds without the constraint of training bias[21–23] and thus addresses the experimentally important question of energetic stability directly. However, its composition-by-composition exploration makes a priori identification of compositions likely to host new structures computationally expensive. While both generative models and CSP aim to propose plausible structures, the generation of chemically reasonable candidates from structural training data by AI is distinct from prediction of the structure corresponding to the ground-state energy at a given composition.

We define a generative task distinct from CSP: learning to propose chemically plausible, structurally diverse chemistry beyond the training distribution. This task has two components: a general plausibility requirement applicable to the scaleable assessment of all generative structural models and a specific requirement to access new structural chemistry. To address this task, we introduce computationally lightweight metrics to assess structural diversity and chemical plausibility based on simple bonding criteria introduced below. We define the local-environment diversity ($M_{LED}$) as an interpretable measure of structural diversity through local chemistry and the compactness $C$ as a proxy for global packing and energetic stability. These metrics are embedded directly into a physics-informed diffusion generative model as conditioning targets and loss components, enabling coupled stability- and diversity-aware generation. The same metrics are also used, together with the data-driven statistical proxy potential (SPP) score[24] as a complementary independent plausibility measure, in a scalable downstream filtering and validation workflow, providing a unified set of tools for generating and selecting viable and diverse candidates for DFT-level energy calculation. Both stability and structural diversity can change upon CSP evaluation of the potential energy surface at generated compositions. We thus propose a role for generative AI as a discovery-oriented front-end that targets physics-based methods for the efficient identification of new structural chemistry from chemical space.

## Results

### Metrics and models for stable and diverse crystal structure generation

To develop tools for the task of proposing chemically plausible, structurally diverse materials beyond the training distribution, we use denoising diffusion, a class of models already demonstrated for structure generation[6–9,25,26], because their iterative refinement process naturally accommodates three-dimensional structural constraints. This approach enables direct incorporation of physical laws and structural constraints into each denoising step[7,9,27]. Integrating such models into the discovery workflow involves three tightly coupled stages: training a condition-aware diffusion model with structural and property signals, activating these learnt conditioned pathways during guided sampling to generate candidate structures, and evaluating both plausibility and diversity. Conditioning is therefore embedded during training rather than applied post-hoc, enabling steered generation in the targeted regions of chemical space. This integration allows the model to internalise chemical and structural criteria that would otherwise only be imposed during validation, targeting the generation of candidates that are more plausible and diverse.

The model presented here, PIGEN, builds on a denoising generative architecture[28] for crystal structure prediction (DiffCSP)[9]. We introduce a measure of stability as a physics-informed loss that constrains the denoising trajectory in PIGEN, incorporate stochastic conditional training with label dropout to enable classifier-free guidance (CFG)[29] and integrate both stability and a further chemically informed structural measure, for local environment diversity, as multi-objective conditioning signals.

Diffusion models generate samples by reversing a stochastic corruption process with a learned denoising neural network[9,28,30]. In PIGEN, a crystal is represented by atom types (**A**), fractional coordinates (**F**) within a unit cell, and lattice parameters (**L**), defining periodicity, similarly to other generative architectures[9]. Atom types are treated as categorical variables, while coordinates and lattice vectors are diffused in continuous space using noise distributions that preserve periodicity and physical constraints. The denoising graph neural network learns to jointly reconstruct (**A, F, L**) from their noised versions.

To incorporate physical plausibility directly into denoising, we introduce compactness (*C*) as a physics-informed loss term:

$$C(\mathbf{A}, \mathbf{F}, \mathbf{L}) \; = V_{\boldsymbol{A}}/V_{\boldsymbol{L}}$$

where $V_{\boldsymbol{A}}$ is the total nominal atomic volume, obtained by summing the volumes of atomic spheres calculated from the tabulated radii of all atoms in the unit cell and $V_{\boldsymbol{L}}$ is the volume of the crystallographic unit cell. Thus, *C* measures how tightly the atoms fill the cell (Methods). While the forward diffusion process adds unconstrained Gaussian noise, the physics-informed *C* loss constrains the reverse process to recover physically plausible structures, effectively projecting intermediate states back onto a chemically reasonable manifold. Compactness thus acts as a lightweight proxy for structural plausibility and stability (Fig. 1a).

For property-aware generation, property labels, such as energy above hull ($E_{hull}$) or Compactness, are introduced alongside structure inputs and a controlled fraction of training steps omit this descriptor (stochastic label dropout). This ensures the model learns both conditional and unconditional denoising trajectories, forming the basis for CFG during sampling (Fig.1b).

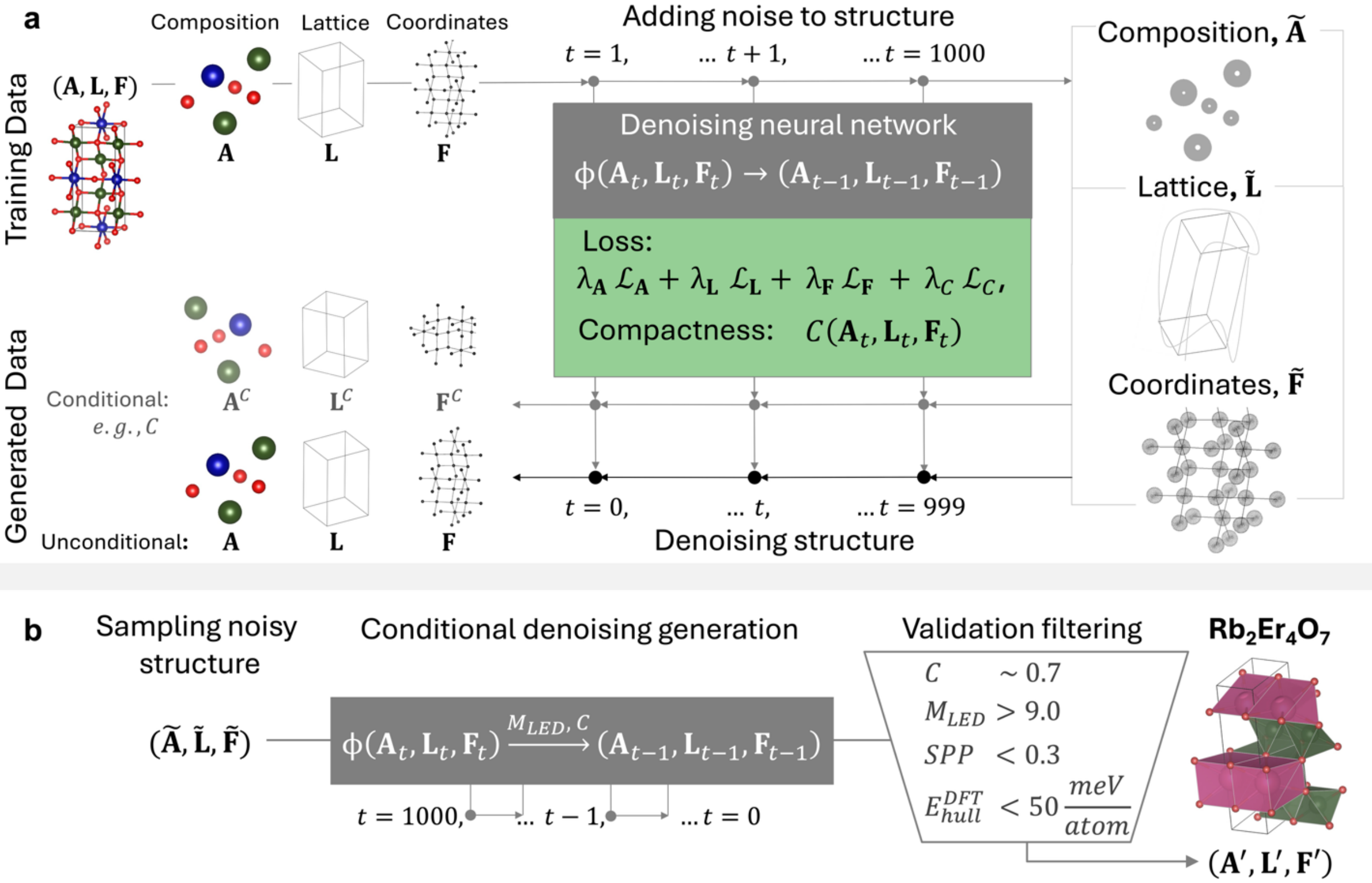

**Figure 1 Training schematics of PIGEN.** The physics-informed denoising neural network for crystal structure generation enforces joint symmetry across composition, lattice, and coordinates. (**a**) Crystal structures are represented as tuples of tensors for Composition, Lattice, and Fractional Atomic Coordinates – $(\mathbf{A}, \mathbf{L}, \mathbf{F})$, which are progressively corrupted during diffusion into $(\tilde{\mathbf{A}}, \tilde{\mathbf{L}}, \tilde{\mathbf{F}})$ and denoised to recover the original configuration. Compactness ($C$) is computed at each diffusion step *t*, informing the denoising model (grey block) training via a compound loss function (green block). Two denoising trajectories are shown: the conditional pathway (light grey), where structural features $(\mathbf{A}^C, \mathbf{L}^C, \mathbf{F}^C)$, are guided by property conditioning signals (for example, compactness $C$), and the unconditional pathway (black), which reconstructs structures solely from noisy inputs. During training, stochastic dropout of conditioning labels enables the model to learn both conditional and unconditional denoising, forming the basis for classifier-free guidance during sampling. (**b**) During generation, classifier-free guidance steers sampling towards structures with plausible compactness and higher chemical and structural diversity ($M_{LED}$); validation includes a chemistry-informed evaluation of bonding through the SPP score[24] and energy filtering $E_{hull}^{DFT} < 50$meV atom$^{-1}$ resulting in realistic and novel structures.

To quantify and guide structural novelty, we introduce a metric for local environment diversity ($M_{LED}$) which guides PIGEN sampling towards diversity-aware generation (Fig. 1b). $M_{LED}$ measures variation in both coordination motifs and chemical environments within a structure (Fig. 2). Each atomic site (Fig. 2, panel 1) is assigned to its nearest coordination polyhedron from a reference set[31,32] (panel 2) and combined with the identities of the neighbouring atoms (panel 3); these descriptors are encoded using corresponding indices from the reference set and atomic numbers (*x*-axis in panels 4-5). Shannon information entropy computed directly from these categorical labels would be insensitive to structural similarities – treating, for example, three different 6-coordinate polyhedra as equally distinct from three polyhedra with 2-, 3-, and 4-coordination, despite clear differences in structural diversity. Applying Gaussian kernels with carefully chosen widths ensures that similar motifs contribute overlapping but distinct signals, transforming sparse categorical observations into smooth distributions that yield a physically meaningful measure of structural diversity (Fig. 2, panels 4-5). Shannon entropies calculated separately for the structural and combined, chemical-structural distributions are summed to give $M_{LED}$, a diversity score that increases with both structural and chemical variation. This provides a practical, chemistry-aware tool for both guiding the diffusion model during generation and evaluating the novelty of its outputs. We also analysed a symmetry-based crystallographic complexity measure[33], which, although useful for quantifying Wyckoff multiplicity and structural complexity, shows a compressed dynamic range for the ≤ 20-atom unit cells in our training data, making it a less responsive signal for guiding diversity during generation than $M_{LED}$ (Extended Data Fig. 1).

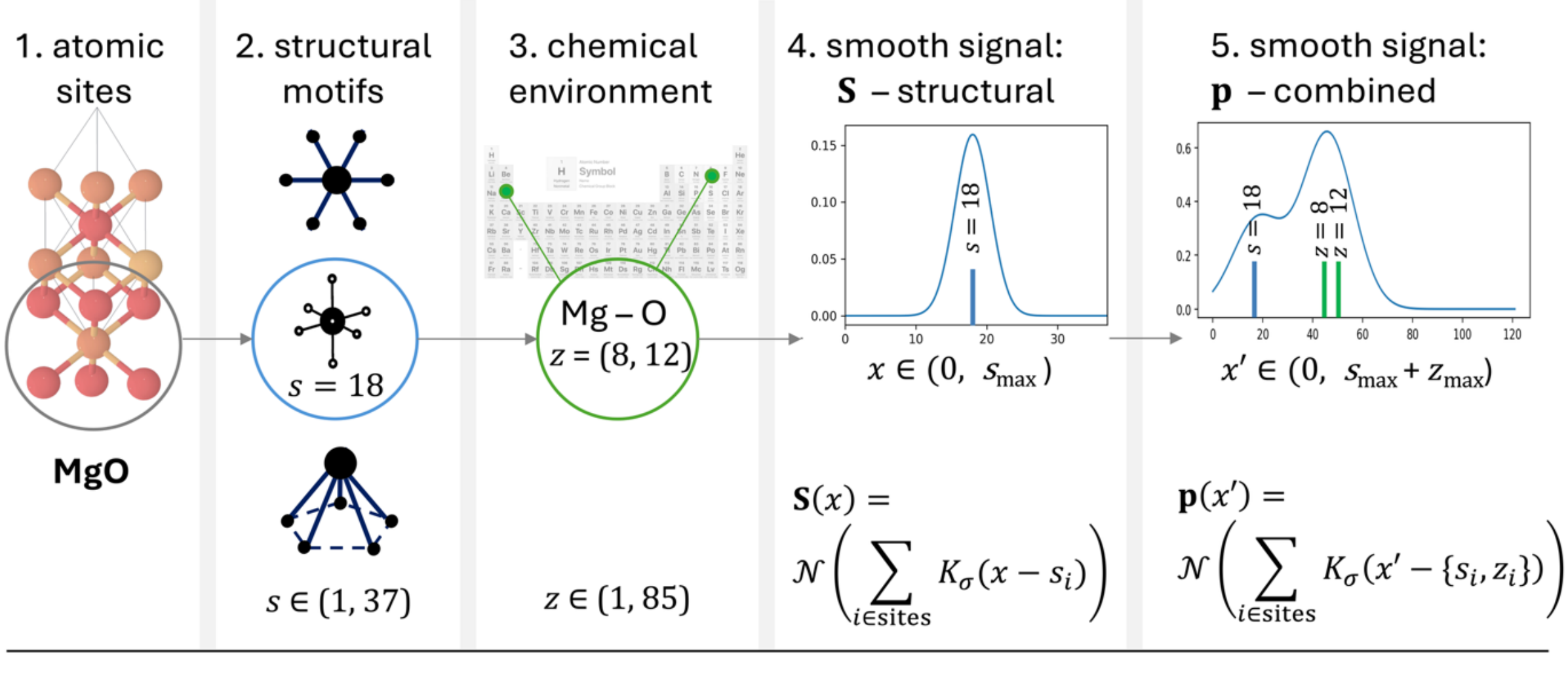

**Figure 2. Derivation of the Local Environment Diversity metric $M_{LED}$.** $M_{LED}$ measures the diversity of local atomic environments in a crystal structure by combining geometric and chemical information, that is, coordination geometry and coordination chemistry. (1) Each atomic site of a crystal structure (example MgO) is examined. (2) Its closest matching coordination polyhedron is identified from a reference library. (3) The chemical environment is defined by the elemental identities of the central and neighbouring atoms, represented by their atomic numbers $z$. (4-5) These discrete structural (4) and chemical (5) values are converted into smooth probability distributions by placing a Gaussian function at each observed value – this smoothing provides robust entropy estimates from sparse data, with the resulting sum of Gaussians defining the probability density shown on the $y$-axis. Summing the Shannon entropies $H$ of these structural and chemical distributions yields $M_{LED}$. High values indicate many distinct coordination motifs and mixed chemical surroundings, whereas a structure with one repeating environment yields a low value. This interpretable metric is used both to guide diffusion-model generation towards diverse outputs and to assess the novelty of generated structures.

### Dataset and training

We train PIGEN models on 607,684 stable crystal structures with up to 20 atoms from the Materials Project (MP)[34] and Alexandria[35] datasets (Alex-MP-20). All element types appearing in Alex-MP-20 were retained, enabling the model to internalise the full chemical diversity presented in explored inorganic chemistry. Each structure is labelled with its $M_{LED}$ and compactness values. Compactness is incorporated directly into the denoising objective so that models learn stability-aware atomic packing reconstruction jointly with atom types, lattice and coordinates. Training and validation show comparable convergence with and without compactness loss (Supplementary Fig. 1-2), indicating that introducing the compactness constraint does not impair training.

During sampling and generation, the number of atoms is allowed to extend up to 30, enabling extrapolation of learned structural motifs to larger unit cells and compositions beyond the size range seen in training. In the property-guided variants discussed in the next section, compactness and $M_{LED}$ are used as conditional signals with partial label dropout, enabling CFG during sampling and multi-objective steering across stability and diversity.

**Property-guided sampling and generation of candidate structures**

In addition to baseline diffusion and unconditional PIGEN models, we train property-conditioned variants to examine how the proposed metrics can guide structure generation across models. To steer generation towards desired regions of chemical space, property-conditioned PIGEN variants employ target values such as compactness, $M_{LED}$, energy above the convex hull, $E_{hull}$, or crystallographic complexity $K$[33] that are incorporated during training with partial label dropout to enable classifier-free guidance (CFG). During sampling, CFG activates these conditional pathways, allowing controlled exploration of structures that satisfy one or more specified objectives (Extended Data Table 1). For brevity, we denote property-conditioned models as PIGEN | $X$, where $X$ is the conditioning target, e.g., PIGEN | $C = 0.7$ when sampling at a specific value, and PIGEN | ($M_{LED} = 9$, $C = 0.7$) for multi-objective conditioning.

Across property targets, we observe clear signatures of how each conditioning objective shapes the generated distribution. Stable materials in the training set cluster around $C \approx 0.7$ (Fig. 3a) based on the observed correlation with $E_{hull}^{DFT}$. Consistent with this, both PIGEN | $E_{hull}$ and PIGEN | $C$ produce similar fractions of structures that pass the different criteria in the validation workflow discussed later (*Validation of plausibility and evaluation of targeted properties*), showing that compactness provides a computationally lightweight proxy for energy-based guidance (Fig 3b). Out-of-distribution (OOD) behaviour can be quantified as the enrichment of generated samples in the low-probability regions of the training distribution.[36,37] The $M_{LED}$ metric highlights the sampling of local-environment space that is rarely represented among known compounds across generative models e.g., 24% of structures generated by (PIGEN | $E_{hull}$) lie above the 99th percentile of the training distribution (Fig. 3c), indicating OOD sampling of local environments according to the definition above. Guiding generation solely by $M_{LED}$ shifts the diversity of generated structures further beyond that of the training set, with 43% (PIGEN | $M_{LED}$) of generated structures above the 99% threshold, emphasising its utility for both identifying and amplifying local structural diversity. It is interesting to note that prototype coverage over ICSD structure types differs markedly across models (Fig. 3d): PIGEN | $E_{hull}$ concentrates 58% of its plausible structures within the 100 most frequent ICSD motifs, with only 42% of structures unmatched by these motifs (Methods). PIGEN | ($M_{LED} = 9$, $C = 0.7$) increases the fraction of plausible structures not matched to top-100 ICSD prototypes to 67%. Although automated prototype matching provides only an approximate, scalar view of structural novelty,[17] this shows that $M_{LED}$-guided sampling can deliberately shift generation away from structural families that are dominant in the training data and towards less explored coordination motifs, aligned with the task of identifying plausible structurally diverse chemical space.

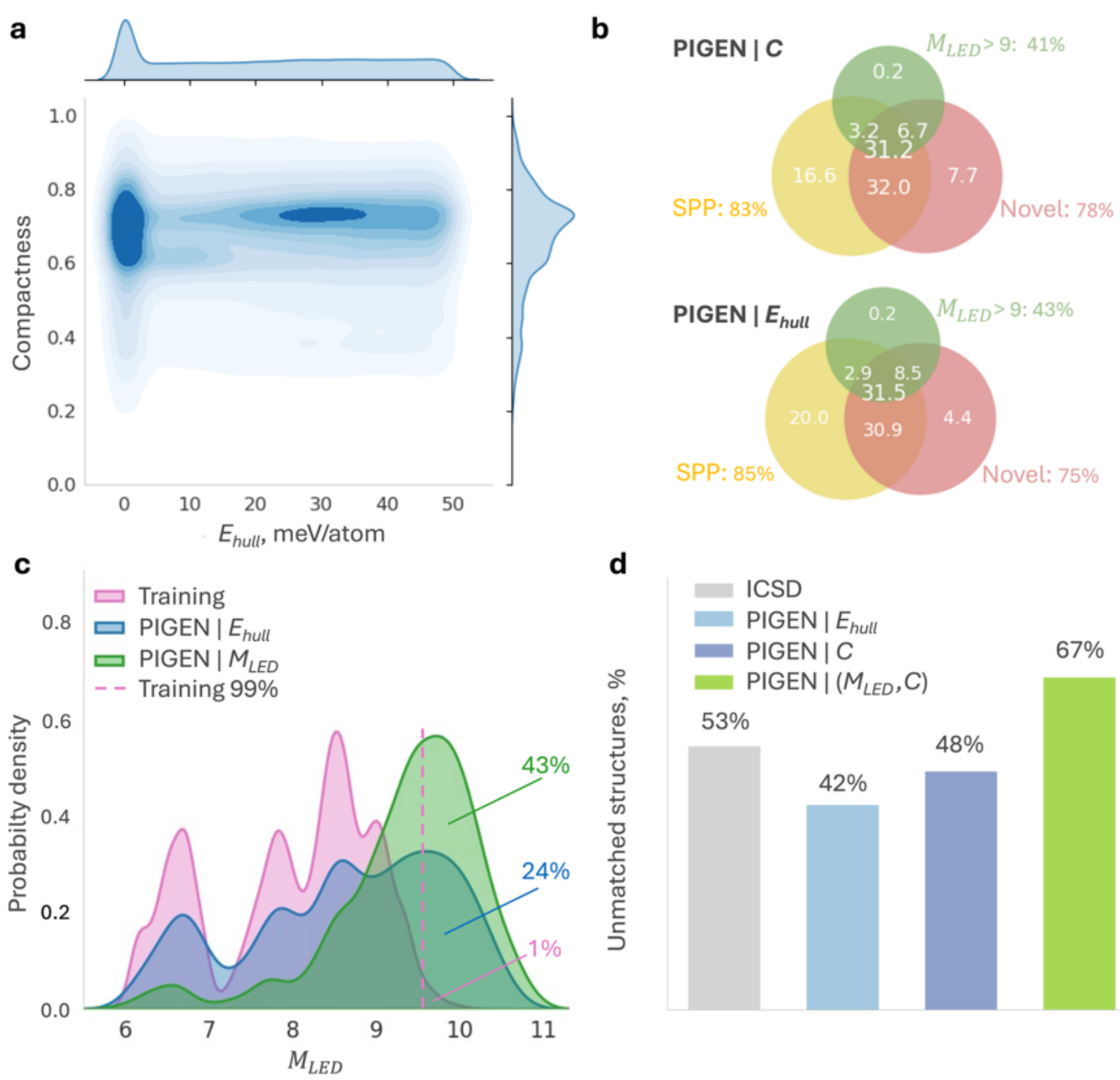

**Figure 3. Results of conditioning and evaluation. (a)** Correlation between compactness $C$ and formation energy $E_{hull}^{DFT}$ in the training set, highlighting compactness as a physically meaningful computationally lightweight proxy descriptor of structural plausibility and stability. **(b)** Venn diagrams comparing conditioning on compactness (top) or energy above hull $E_{hull}$ (bottom), showing that both targets yield a nearly equivalent fraction of structures satisfying SPP validity, compositional novelty, and $M_{LED}$. **(c)** Distribution of $M_{LED}$ values for the training structures (pink) and structures generated by PIGEN | $E_{hull}$ (blue) and PIGEN | $M_{LED}$ (green). $M_{LED}$ conditioning enriches sampling in the low-probability tail of the training distribution, corresponding to chemically and structurally diverse local environments that are rarely realised in reported crystal structures. **(d)** Fraction of plausible generated structures – filtered for SPP validity, compactness, and compositional novelty – which are unmatched to the top-100 most frequent ICSD[38] structural prototypes. For PIGEN | $E_{hull}$, 42% ($n$ = 2,957) of plausible structures are unmatched, compared to 67% ($n$ = 5,189) for PIGEN | ($M_{LED}$, $C$). This indicates that $M_{LED}$-guided sampling shifts generation away from structural frameworks that are prevalent in the training data. 53% of structures in ICSD are unmatched by the top-100 prototypes.

## Validation of plausibility and evaluation of targeted properties

A key requirement for our task is to identify which generated structures are both chemically plausible and structurally novel before committing to expensive DFT relaxations. To maintain generality across chemical space, we do not apply charge-balancing constraints, which are only appropriate for ionic materials and would bias sampling against intermetallic and covalent frameworks. Simple heuristics based on 0.5 Å interatomic distance cutoffs[4–11] classify the vast majority of generated structures (≈ 99%) as valid, but fail to detect unrealistic local chemical bonding patterns or implausible packings. We therefore develop a chemically grounded, scalable validation workflow that filters generated candidates using global and local structural validity metrics together with compositional novelty to assess chemical plausibility and redundancy and then applies structural diversity as a lightweight filter for the task of identifying new structures. These four filtering stages (***A***–***D***) each remove a distinct fraction of unrealistic, redundant or off-task candidates before final DFT assessment of stability (***E***, Fig. 4).

***A.*** *Structural validity* (Fig. 4a). We use the Statistical Proxy Potentials (SPP) criterion[24] which compares interatomic distances against element-specific distributions derived from ICSD[38] (Methods). This local structural measure rejects ~24% of structures that pass simple distance cutoffs across all models, ensuring chemically reasonable bonding environments. For example, in the generated BiS structure, all pair distances exceed 1 Å, yet Bi–Bi, Bi–S and S–S separations deviate from ICSD bonding statistics, yielding a high SPP score and indicating chemically unrealistic geometry.
***B***. *Compositional novelty* (Fig. 4b). This metric removes candidates whose elemental combinations appear in the training set. For example, $RbErO_2$ would be discarded, whereas $Rb_2Er_4O_7$ is retained as a novel composition. Applied alone, this filter disqualifies ~26% of generated structures across models.
***C***. *Compactness* (Fig. 4c). Compactness acts as a global plausibility measure. Structures with very low $C < 0.3$ show expanded lattices with large, chemically unrealistic voids. For example, the generated $FeAs_8O_{19}$ structure shows extensive void regions[39], whereas the $Rb_2Er_4O_7$ structure in the lower panel illustrates a compact chemically reasonable packing. We retain structures with $0.55 < C < 0.85$ reflecting the correlation with stability in the training data noted earlier. This window could be tuned to target other classes of materials, such as porous materials. On average, 18% of generated candidates that pass local structural validity assessed by the SPP score are removed by this global $C$ filter.
***D***. *Local environment diversity* ($M_{LED}$) (Fig. 4d). $M_{LED}$ quantifies diversity in generated coordination motifs and chemical environments beyond compositional changes. Increasing $M_{LED}$ values mark transitions between dominant structural families (e.g., fcc, NaCl, perovskite, spinel, etc.) and correlate with a progressive departure from the most frequent ICSD prototypes and into the "other" category, as illustrated in Fig 4d for the PIGEN | $E_{hull}$ model. A threshold of $M_{LED} > 9$ is used to select high-diversity structures, capturing the upper 15% of the training distribution, and serving as a practical cutoff for prioritising compositions likely to afford novel structural chemistry for further evaluation with more expensive downstream methods. Details of prototype matching are provided in Methods.

Sequential filtering (***A–D***) (Fig. 4, Extended Data Fig. 2) progressively distils raw generative outputs – initially dominated by implausible or redundant structures – into a small subset of chemically feasible and structurally diverse candidates using chemistry-grounded metrics. This validation workflow is based on a model-agnostic post-generation filter evaluating bonding environments (SPP), global packing (compactness) and compositional novelty for reliable down-selection at scale from any model for crystal structure generation.

Once such physically plausible candidates are established, task-specific criteria can be applied. Here, high diversity ($M_{LED}$) is used to prioritise compositions that deviate from well-represented structural families and drive exploration of less-sampled coordination motifs. Stage ***E*** then applies DFT relaxation and convex-hull evaluation, selecting candidates within 50 meV $atom^{-1}$ of the convex hull as energetically viable using the hull as a standard measure of stability relative to competing phases and 50 meV $atom^{-1}$ as a conservative metastability window[40]. Because these calculations are computationally demanding, the lightweight filters ***A–D*** focus them on promising and diverse candidates. The small fraction of outcomes within 50 meV $atom^{-1}$ of the convex hull reflects the trade-off between structural novelty and stability introduced by diversity-focused filtering[38]. This highlights the complementary roles of generative models as exploratory tools for identifying promising regions of chemical space, and physics methods as the final validation of chemical and energetic plausibility.

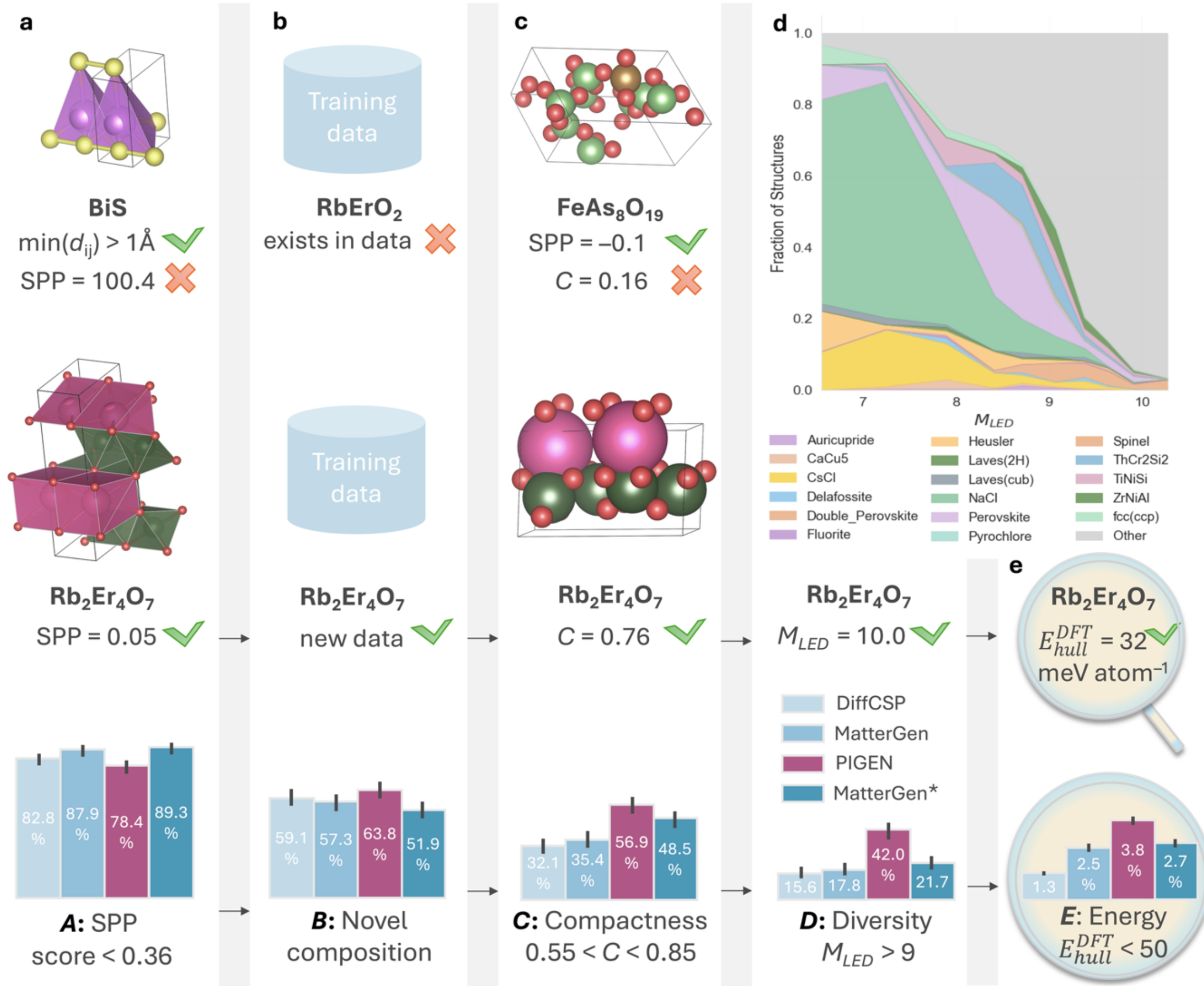

**Figure 4. Validation and evaluation of generated crystal structures.** Generative model outputs of 10,000 structures per batch are filtered through a multi-stage protocol to identify chemically plausible and novel candidates. **(a)** Structural validity is assessed using Statistical Proxy Potentials (SPP)[24], which compare interatomic distances against element-specific interatomic distance distributions derived from ICSD; **(b)** Compositional novelty retains structures with element combinations absent from the training set. **(c)** Compactness enforces physically informed plausibility, removing structures with unrealistic global packings. **(d)** Local environment diversity ($M_{LED}$) quantifies structural novelty beyond composition and marks transitions between dominant structural families (e.g., fcc, NaCl, perovskite, spinel). Higher $M_{LED}$ enriches the "other" category – frameworks not matching any of the top 20 most frequent ICSD prototypes shown here (selected for clarity of visualisation) **(e)** DFT evaluation of generated structures to identify candidates within 50 meV atom$^{-1}$ of the convex hull. The histograms across stages show the progressive refinement of the fractions of structures generated by various models: DiffCSP[9], MatterGen[7], PIGEN | ($C = 0.7$, $M_{LED} = 9$) and MatterGen | ($C = 0.7$, $M_{LED} = 9$) (MatterGen* in the legend). The scaleable, descriptor-guided filtering pipeline enables efficient triage of raw generative outputs, ensuring that only chemically plausible and high-diversity structures proceed to costly DFT evaluation.

## Generating diverse and plausible crystal structures

To benchmark PIGEN, we compared it with several baseline generators under the same validation workflow (***A***–***E***). The baselines include: pseudo-random structure generation, analogous to random atom placement within the unit cell (Methods), included to illustrate the absence of learned chemical or structural priors and resulting in essentially no valid outputs; DiffCSP[9] retrained on Alex-MP-20 without the physics-informed loss or classifier-free guidance (CFG), representing an unconditioned diffusion baseline; the published MatterGen model[7], serving as a state-of-the-art reference for crystal generation; and MatterGen fine-tuned for multi-objective optimisation on Compactness and $M_{LED}$, used to assess the transferability of the proposed metrics to alternative architectures.

In contrast to MatterGen, which introduces property steering via adapter-based fine-tuning of a small set of layers after pretraining, PIGEN incorporates Compactness and $M_{LED}$ as conditioning signals directly during full model diffusion training (with partial label dropout) so that all network parameters learn the property-structure relationship, resulting in tighter coupling between property steering and structure generation. As summarised in Table 1, conditioning on the Compactness and $M_{LED}$ increases the fraction of chemically plausible and structurally novel candidates across models (column *D*), improves the proportion of structures at low energies above the convex hull (column *E*), and accesses higher local-environment diversity $M_{LED}$ (column *F*). For the specific task of generating stable yet structurally diverse frameworks – the focus for which PIGEN was designed – PIGEN | ($M_{LED}$, *C*) achieves the strongest overall performance: 42.0 ± 0.5% valid, diverse candidates, compared with 21.7 ± 0.5% for the best MatterGen variant and 17.8 ± 0.5% for MatterGen, with all pairwise differences highly significant (two-proportion *z*-test, $p \ll 10^{-12}$; see Supplementary Note 1). For candidates within 50 meV atom$^{-1}$ of the convex hull (column *E*), PIGEN | ($M_{LED}$, *C*) similarly outperforms all baselines in the fraction of valid, low-energy structures, with differences again highly significant (Supplementary Note 1). Additional variants, including models conditioned on single properties ($E_{hull}$, *C*, $M_{LED}$, *K*) and their combinations, are benchmarked in Extended Data Table 2 to provide a comprehensive comparison across all conditioning targets.

**Table 1. Performance of generative models on a 10,000-structure batch.** Column *D* reports the percentage of generated structures that satisfy the validity and diversity criteria of Fig. 4a-d. Column *E* adds energy filters of increasing stringency, with cumulative standard error after ***A–E*** filters ranging from 0.1–0.3% (Extended Data Fig. 3); individual SEs are omitted from the table cells in Column *E* for clarity. Column *F* gives the maximum $M_{LED}$ diversity among the Column *E* structures with $E_{hull}^{DFT}$ < 50 meV atom$^{-1}$. *Notation*: MODEL | ($X = x$, $Y = y$) indicates the corresponding MODEL is multi-conditioned on both *X* and *Y* and sampled at values $X = x$, $Y = y$. Bold font indicates best performance (higher values), underscored font – second best.

| | ***D*** | ***E*** | ***F*** |
|---|---|---|---|
| **Model** | Filters ***A–D***,<br>Fig. 4 a & b & c & d<br>% | ***D*** & Stable:<br>$E_{hull}^{DFT}$ < 100 / 50 / 35 / 25<br>meV atom$^{-1}$, % | max($M_{LED}$)<br>for $E_{hull}^{DFT}$ < 50<br>meV atom$^{-1}$ |
| Pseudo-random (Methods) | 0.03 | 0 / 0 / 0 / 0 | - |
| DiffCSP[9] | 15.6 ± 0.5 | 4.0 / 1.3 / 0.7 / 0.5 | 10.7 |
| MatterGen[7] | 17.8 ± 0.5 | 5.1 / 2.5 / 1.6 / 1.0 | 10.7 |
| MatterGen \| ($M_{LED}$ = 9, *C* = 0.7) | 21.7 ± 0.5 | 6.7 / 2.7 / 1.8 / 1.2 | 10.7 |
| **PIGEN \| ($M_{LED}$ = 9, *C* = 0.7)** | **42.0 ± 0.5** | **8.7 / 3.8 / 2.3 / 1.5** | **10.9** |

While these results among novel plausible candidates reflect proximity to the convex hull and thus to 0 K thermodynamic stability, they do not guarantee that the generated structures correspond to the true energetic ground states at each identified composition. To illustrate this, we selected two charge-balanced ionic compositions from among the 130 filtered structures generated by PIGEN | ($M_{LED}$, *C*) that both lie close to the convex hull ($E_{hull}^{DFT}$< 50 meV atom$^{-1}$) and were identified as not matching any ICSD prototypes, for detailed exploration of their potential-energy surfaces using crystal structure prediction (Methods, Fig. 5). For these two cases, we also analyse the distributions of local structural environments, as defined in Fig. 2 panel 4, comparing the training data with the generated and CSP-relaxed structures (Fig. 5, insets). The generated structure of $LiNb_7N_8$ (Fig. 5a) is initially unstable ($E_{hull}^{DFT}$ > 0 meV atom$^{-1}$); after global relaxation with CSP (Methods), it transforms into the known $W_2C$(hP3) prototype, becoming energetically stable ($E_{hull}^{DFT} < 0$ meV atom$^{-1}$). In contrast, the initially unstable generated structure of $Rb_2Er_4O_7$ relaxes during CSP optimisation into a distinct, previously unreported stable ($E_{hull}^{DFT} < 0$ meV atom$^{-1}$) framework absent from all ICSD prototypes (Fig. 5b). The corresponding CIF files for these four structures are provided in Supplementary Data 1. Consistent with this, the $M_{LED}$-based environment distributions in Fig. 5b show pronounced deviations from the training data, reflecting the emergence of new coordination motifs in the $Rb_2Er_4O_7$ framework, whereas those in the CSP-derived structure in Fig. 5a lie within the training data.

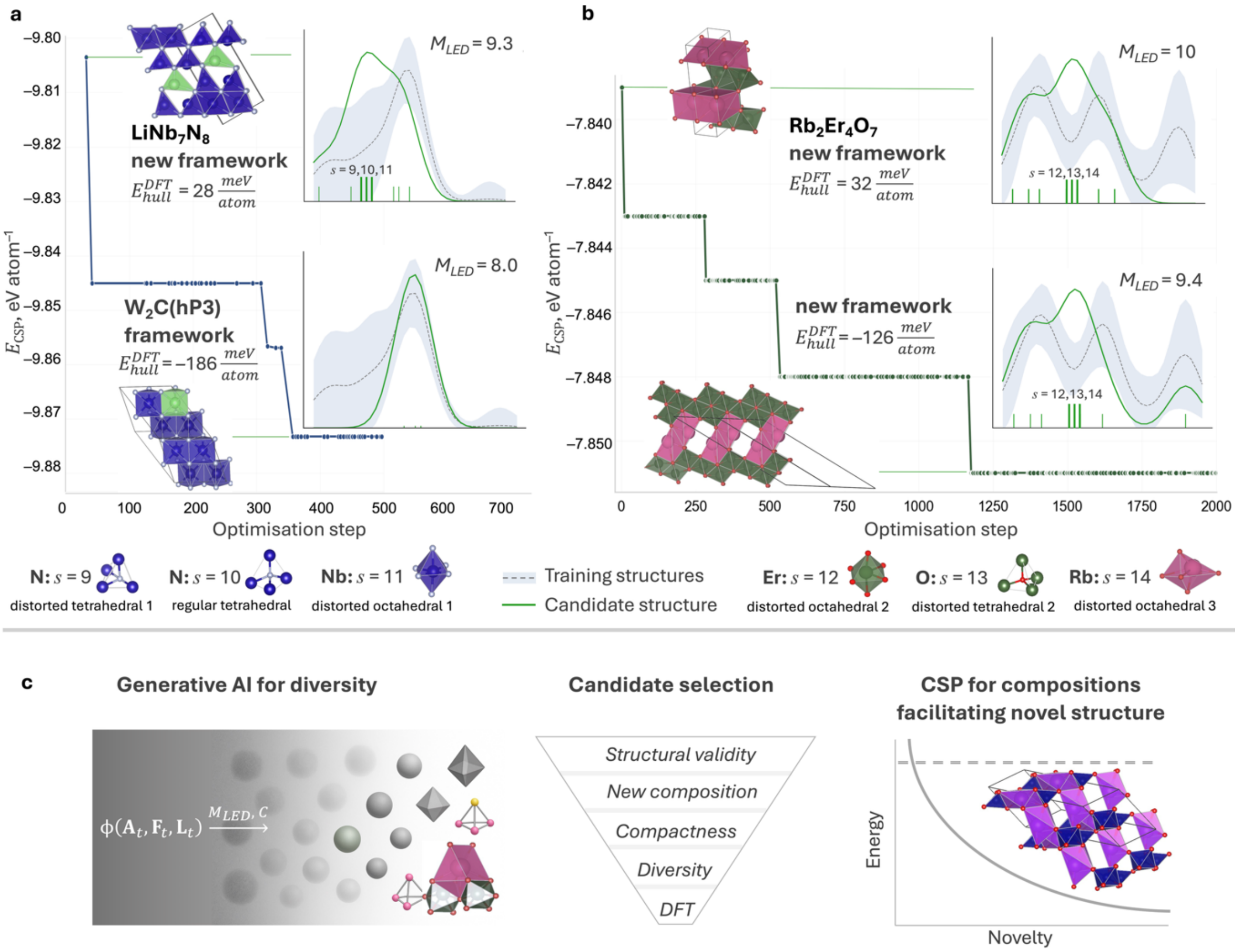

**Figure 5. Global optimisation of PIGEN | ($M_{LED}$, $C$)-generated structures identified as low-energy novel frameworks. a** Basin-hopping optimisation with the crystal structure prediction (CSP) code FUSE[42] of $LiNb_7N_8$. The CSP energy trace ($E_{CSP}$) shows successive reductions and plateaux as new minima are discovered (Methods); the final structure relaxes to the known $W_2C$ prototype and lands on the convex hull. **b** Equivalent optimisation for $Rb_2Er_4O_7$, which attains a convex hull energy structure while reforming into a novel framework not matching any entry in the ICSD, illustrating stabilisation of a potentially novel motif when optimisation starts from a high-diversity ($M_{LED} > 9$) candidate. Insets in **a** and **b** show the distributions of local structural environments, as defined in Fig. 2 (panel 4), in the Li–Nb–N and Rb–Er–O systems for the training data (dashed line; shaded band indicates the standard error; Methods) and for the corresponding candidate structures before and after CSP (green line). Indices of prominent structural motifs that deviate from the training data, and thus yield high $M_{LED}$ values, are marked with tall ticks along the $x$-axis and corresponding motifs are visualised below. **c** A practical workflow for discovery-oriented exploration of composition-structure space. Generative models guided by diversity and viability metrics identify compositions favouring previously unreported structural motifs. After filtering for chemical plausibility and high local environment diversity $M_{LED}$, these compositions serve as informed starting points for CSP, which then explores their potential energy surfaces. This integrated approach enriches exploration by guiding CSP towards previously uncharted compositions that favour new structural chemistry for experimental study.

These examples emphasise that full global structural optimisation via CSP remains essential even after chemically informed filtering, as it can reveal thermodynamically viable frameworks, both known and unknown, that may be experimentally accessible, overturning both stability and novelty outcomes. The DFT energy reductions of 214 and 158 meV atom$^{-1}$ respectively in the two cases reinforce this point, also noting that the CSP global minimum here is itself from a heuristic and thus not guaranteed[43].

## Discussion

Scalable validation of candidate plausibility is critical for assessment of any generative crystal structure model, as it enables meaningful model comparison and efficient prioritisation for expensive but essential DFT evaluation. By integrating compactness and SPP score as complementary proxies for global packing efficiency and local bonding correlated with structural stability, we establish a chemically-grounded framework for high-throughput plausibility assessment of generative model outputs. For the task considered here – proposing chemically plausible, structurally diverse frameworks beyond the training distribution – we incorporate compactness directly into the generative objective to promote stability and introduce $M_{LED}$ to quantify and control diversity of geometric and chemical local environment. The validation workflow applies the task-specific measures after plausibility is established, focussing the expensive DFT calculations onto the appropriate chemistries.

Introducing conditioning on both compactness and $M_{LED}$ shifts sampling towards regions of structural chemistry that are underrepresented in existing databases by expanding coordination geometry and coordination chemistry, increasing the breadth of local environments and geometric motifs accessible to the model. This conditioning systematically improves the plausibility and diversity of generated structures across models, including strong baselines, by shifting sampling away from motifs that dominate the training data while maintaining stability. The combined conditioning and filtering thus enables scalable evaluation and DFT targeting of candidate compositions favouring stable structures with out-of-distribution local environments.

Physics-informed generative models do not perform crystal structure prediction, but can propose chemically plausible, diversity-targeted candidate compositions that guide downstream optimisation. This coupling offers a scaleable route to identify realistic yet novel candidate structures, exemplified here by the $Rb_2Er_4O_7$ candidate that reached the DFT-confirmed convex hull at 0 K after CSP refinement. More broadly, generative AI provides a fast, information-efficient front end for proposing compositions and structures targeted on novelty and plausibility. DFT then CSP establish the actual physical viability and the associated novelty of appropriately filtered candidates from the generative models. This suggests a synergistic workflow in which generative AI guides CSP towards uncharted regions of chemical space rich in new structural chemistry (Fig. 5c).

## Methods

### Diffusion-Based Crystal Structure Generation

Crystal structures are generated via Denoising Diffusion Probabilistic Models (DDPM). The forward process adds Gaussian noise to data $x_0$, and a neural network $\phi(x_t, t)$ learns to estimate the noise to reverse the process.

The model predicts noise from atom types **A**, lattice parameters **L**, and fractional coordinates **F** jointly, with the objective

$$\mathcal{L}_{base} = \lambda_{\mathbf{A}}\, \mathcal{L}_{\mathbf{A}} + \lambda_{\mathbf{F}}\, \mathcal{L}_{\mathbf{F}} + \lambda_{\mathbf{L}}\, \mathcal{L}_{\mathbf{L}} \tag{1}$$

where $\lambda_i, \boldsymbol{i} \in \{\mathbf{A}, \mathbf{L}, \mathbf{F}\}$ are the weighting factors to the corresponding loss components. More details on the DDPM equations, marginal distributions and noise predictions are provided in Supplementary Information (SI).

### Physics-informed equivariant diffusion

We introduce a physics-informed loss term to enforce chemically plausible structure compactness:

$$C = \frac{4}{3}\pi \frac{\sum_a r_a^3}{V_L}, \tag{2}$$

where $r_a$ are standard atomic radii[44], giving the total atomic volume in the unit cell, and $V_L$ is the unit cell volume defined by the lattice parameters $L$. We ignore potential overlap of atomic spheres for simplicity. Let $C(A_t, F_t, L_t)$ denote predicted scalar value compactness at timestep $t$ and $C_0$ the reference, then the loss function at each diffusion step $t$ is given by:

$$\mathcal{L}_t = \mathcal{L}_{base} + \lambda_C |C(A_t, F_t, L_t) - C_0|^2 \tag{3}$$

where $\mathcal{L}_{base}$ is the base loss function Eq. (1) and $\lambda_C$ is a weighting factor. Weighting factors $\lambda_A$, $\lambda_F, \lambda_L$ for the base loss and $\lambda_C$ were selected to rescale individual loss terms, ensuring balanced gradient magnitudes and stable joint optimisation across all objectives. The compactness term in Eq. (3) ensures predicted structures remain chemically and physically realistic throughout the diffusion process. We introduce a compactness-preserving property (Proposition 1 in SI) that guarantees equivariance of predicted structures under permutations, translations, and lattice rotations. Equivariance ensures that applying these symmetry operations to the input produces correspondingly transformed output, preserving physical consistency. The proof of Proposition 1 is given in SI. We note that while the forward diffusion process employs standard Gaussian noise addition without physical constraints, this is intentional: the physics-informed loss term during training teaches the model to denoise arbitrary corrupted structures by projecting them back onto the physically plausible manifold. This approach allows the model to handle diverse initial conditions while ensuring physical consistency in generated structures. Alternative approaches enforcing constraints in both forward and reverse processes (e.g., constrained diffusion on Riemannian manifolds) are possible but significantly increase computational complexity.

### Local Environment Diversity, $M_{LED}$

**Formal definition.**

$M_{LED}$ is defined as the sum of two Shannon entropy terms:

$$M_{LED} = H(\mathbf{S}) + H(\mathbf{p}), \tag{4}$$

where

$$H(\mathbf{S}) = -\textstyle\sum_m \mathbf{S}(x) \log \mathbf{S}(x), \tag{5}$$

$$H(\mathbf{p}) = -\textstyle\sum_m \mathbf{p}(x') \log \mathbf{p}(x'), \tag{6}$$

Here, $\mathbf{S}(x)$ is a normalised Gaussian mixture capturing the frequency distribution of local coordination motif types $x \in \{1, \ldots, s_{max}\}$ observed across atomic sites, following a set of 37 common coordination polyhedra[31,32]. $\mathbf{p}(x')$ is a normalised Gaussian mixture over all chemically and structurally active indices $x' \in \{1, \ldots, s_{max}+z_{max}\}$ spanning the concatenated index sets for both structural motifs $s$ and atomic numbers $z$.

**Construction of the distributions.**

*Local assignment.* For each atomic site in a crystal structure, represented as the symmetry-reduced primitive cell with a unique lattice orientation via pymatgen[44], we identify (i) the closest-matching coordination polyhedron from a set of common 37 motifs[31], and (ii) the chemical environment defined by the element type of the central atom and its neighbours, represented by atomic number (up to $z = 85$). These assignments are encoded as multi-hot vectors: $x$ for motif types and $x'$ for concatenated motifs and atomic-species indices, with ones at the active indices and with zeros elsewhere.
*Continuous representation.* Active indices $i$ for each atomic site are projected onto a one-dimensional axis via a Gaussian kernel of fixed variance, producing a smooth per-site signal over the discrete index domain.
*Global aggregation.* Summing the per-site signals over all sites in the structure yields the total distribution $\mathbf{p}(x')$ which reflects the frequency and overlap of both structural motifs and chemical types. Similarly, summing continuous representations of motif counts across all sites gives $\mathbf{S}(x)$. Both distributions are then normalised to integrate to one.
*Entropy calculation.* Finally, we compute Shannon entropies of these normalised distributions $H(\mathbf{S})$ and $H(\mathbf{p})$ and define $M_{LED} = H(\mathbf{S}) + H(\mathbf{p})$.

**Interpretation.**

$M_{LED}$ increases with the variety of coordination environments and elemental site types present in the crystal. Because it captures both geometric and chemical heterogeneity in a single continuous quantity, $M_{LED}$ serves as an interpretable proxy for local-structure novelty and is used to guide and evaluate generative crystal-structure sampling throughout this work.

**Computation of training-set environment distributions for Fig. 5.**

For the Li–Nb–N and Rb–Er–O case studies in Fig. 5, we restricted the training set to structures containing all three elements Li, Nb and N (60 structures) and Rb, Er and O (89 structures), respectively. For each such training structure we constructed the distribution of local structural environments $\mathbf{S}(x)$ as defined above. The dashed curves in Fig. 5 show the mean of these per-structure distributions across the corresponding subset of training structures, and the shaded bands indicate the standard error around this mean.

### Conditional generation via classifier-free guidance

We adopt a classifier-free guidance (CFG) scheme[29] to enable property-conditioned crystal generation without training an auxiliary predictor, in contrast to the approach in DiffCSP. In CFG, the model predicts both conditional and unconditional noise $\hat{\epsilon}$, which are linearly combined during inference into $\hat{\epsilon}^*$:

$$\hat{\epsilon}^*(z_t, t \mid \mathbf{q}) = (1 + g)\, \hat{\epsilon}(z_t, t \mid \mathbf{q}) - g\, \hat{\epsilon}(z_t, t), \tag{7}$$

where $g$ modulates guidance strength. Conditioning information $\mathbf{q}$, representing properties such as $C$ or $M_{LED}$, is randomly dropped with probability $p_{drop}$, which is set to balance unconditional and conditional updates during training. Aggregating noises for $(\mathbf{A}, \mathbf{L}, \mathbf{F})$ and Compactness modulates the structure towards the target conditions. More details on conditional embedding, drop probability, and aggregation over $(\mathbf{A}, \mathbf{L}, \mathbf{F})$ are provided in SI.

### Statistical Proxy Potential (SPP) Filtering

To rapidly assess chemical plausibility of generated crystal structures we adopt the Statistical Proxy Potential (SPP) method of Ref.[24].
SPP provides an element-resolved score of structural realism by comparing all interatomic separations for considered elements in a candidate structure with probability distributions derived from experimentally reported crystals in the Inorganic Crystal Structure Database (ICSD)[38].

*Construction of the potential*
For every element pair $(i, j)$, ICSD interatomic distances are binned to form empirical probability distributions $P_{ij}(r)$. From these distributions an effective pairwise potential,

$$U_{ij}(r) \; = \; -\, k_B T \ln P_{ij}(r)$$

is defined, yielding a purely data-driven statistical proxy potential that captures the range of separations observed in stable compounds.

*Scoring a candidate structure*
Given a generated crystal, all interatomic distances $r_{ab}$ (including periodic images) are evaluated against the corresponding $U_{ij}(r)$. The SPP score is the mean of these pairwise potentials,

$$\mathrm{SPP} = \frac{1}{N_{\mathrm{pairs}}} \sum_{a<b} U_{A_a A_b}(r_{ab})$$

where $A_a$ denotes the species of atom $a$.
Lower scores indicate closer agreement with the experimental distributions.

*Application*
We use SPP as a lightweight chemistry-aware filter before computationally expensive density-functional theory (DFT) calculations.
Structures with scores exceeding the 95th-percentile threshold determined from ICSD (SPP score > 0.362 in our dataset) are discarded as chemically implausible. This element-specific screening removes grossly unrealistic geometries – such as under-bonded clusters or compressed contacts – while preserving candidates likely to survive full electronic-structure optimisation.

### Crystal structure prototype matching

Generated structures were compared to 9523 known structural prototypes reported and labeled in ICSD as distinct structure types, using the `StructureMatcher` class in pymatgen[44]. To capture framework equivalence independent of chemical decoration, we applied the `FrameworkMatcher` comparator, which anonymises atomic species and compares lattice geometry and connectivity with a tolerance *stol*=0.3. Each candidate was transformed to a standardised primitive cell before matching. Structures identified as equivalent under this species-agnostic comparison were classified as belonging to an existing prototype.

For visual clarity, Fig. 4d illustrates the distribution of local-environment diversity ($M_{LED}$) using only the 20 most frequent ICSD prototypes. Large-scale cross-model comparisons (Fig. 3d and Extended Data Fig. 3b,c) use the top 100 prototypes, evaluated across batches of 10,000 generated structures. For the assessment of scalar structural novelty in this way of the structures in the Fig. 5 the full ICSD prototype set was used for exact matching.

### Crystal structure prediction calculations with FUSE

CSP calculations for compositions in Fig. 5 were performed using FUSE v2.05[42] with reinforcement learning-controlled basin-hopping.[45] During optimisation, energies and structural relaxations used the CHGNet machine-learnt potential[46] with forces converged to below 0.05 eV $Å^{-1}$. Post optimisation, any structure which contains interatomic distances below 1 Å is automatically rejected during the CSP search. For both compositions in Fig. 5, FUSE was restricted to the same number of atoms proposed by the generative model. FUSE initialises the basin hopping search using an inbuilt structure generator to build a pool of structures which it can pull from. For the calculations in Fig. 5, the structures from the generative model were added into this pool, ensuring that the generative model structure was considered during the global optimisation.

FUSE uses a basin hopping routine which measures how many structures have been generated since the current global minimum was located. This value is used to determine when to stop the structural search. For this work, we used a value of 1,500, as is a typical use case. While this means that a minimum of 1,500 structures need to be generated by the basin-hopping, there is no fixed upper limit. This resulted in 2,189 and 2,852 structures for $LiNb_7N_8$ and $Rb_2Er_4O_7$ respectivly.

### DFT energy calculations

Thermodynamic stability was evaluated using single-point DFT total energies computed with VASP-6.2.1[47]. All structures were pre-relaxed using CHGNet, after which single-point DFT calculations were performed on the CHGNet-relaxed geometries to obtain a high-fidelity energies. The same CHGNet → DFT protocol was applied to all reference compounds in the training database to ensure energy comparability across candidates and references. DFT calculations employed the projector-augmented wave formalism with the PBE exchange–correlation functional, a 520 eV plane-wave cutoff, and Γ-centered $k$-point grids generated with a reciprocal resolution $\Delta k_i = 1/25$ $Å^{-1}$ along each lattice direction $i$. Electronic convergence was set to $10^{-8}$ eV, and no further ionic relaxation was performed at this stage. This general protocol was used for the reference set and batch evaluations. For the final evaluation of structures presented in Fig. 5, a CHGNet → DFT protocol was retained, but the DFT calculations for all structures in this phase field were performed using VASP-6.5[47] with 600 eV plane-wave cutoff, electronic convergence $10^{-6}$ eV and the automatic Γ-centered $k$-point generation with KSPACING = 0.3 $Å^{-1}$. The resulting DFT total energies were then used to compute the energy above the convex hull ($E_{hull}^{DFT}$) by comparing each candidate's energy to the set of competing phases in the same phase field in the training data.

### Random structure generation

For our baseline experiment, generating random structures (Table 1 "Pseudo-random"), the Ab-Initio Random Structure Search (AIRSS) code was used[18]. The input file for AIRSS was configured such that the minimum separation between atoms was 1 Å, structures were permitted to contain between 1 and 20 atoms, and it was given a list of the unique elements which appeared in our model training data. We emphasise that this is not a typical use case for AIRSS as a structure prediction tool. Instead, we used its capability to reliably generate a large population of pseudo-random structures to provide a baseline for comparison to the models in this work.

### Statistical analysis

For all reported proportions in the main text, we use Wilson score 95% confidence intervals[48]. Differences in proportions between models (for example, in Table 1) were assessed using two-proportion $z$-tests on binomial counts. In Supplementary Information, risk ratios are reported with log-scale confidence intervals, and odds ratios with standard large-sample approximations. Additional analyses of binary validation pass/fail metrics shown in Table 1 and Extended Data Table 1, 2 use the standard binomial standard error estimator and cumulative standard error[4749] where multiple metrics are applied sequentially (Supplementary Figure 5).

## Extended Data

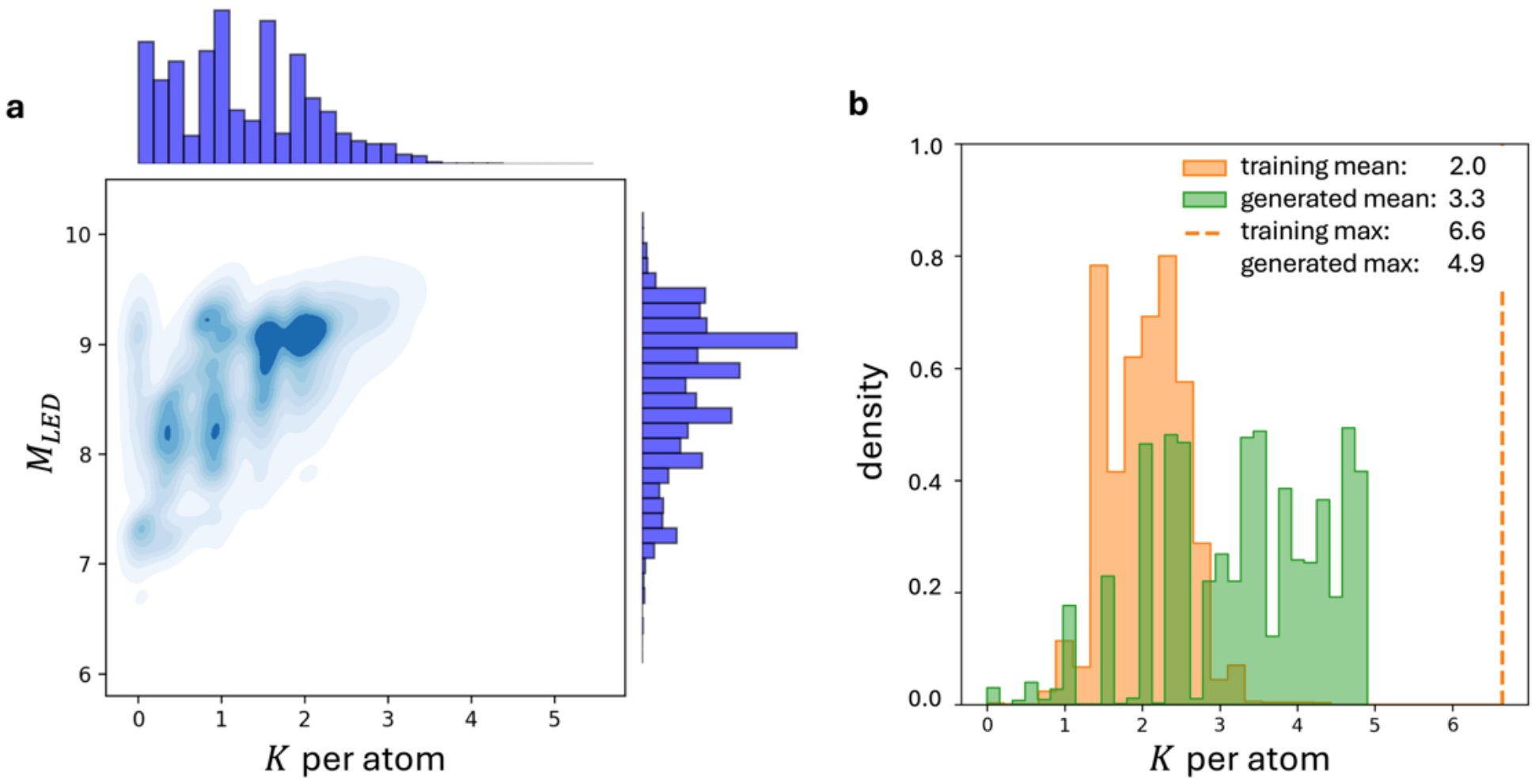

**Extended Data Figure 1. Comparison of crystallographic complexity[33] and diversity metrics.** **a** Density plot comparing an established crystallographic complexity measure per atom ($K$/atom, Ref. [33]) with the local-environment diversity metric ($M_{LED}$) for crystal structures in the ICSD with unit cells limited to ≤ 20 atoms as in the training data. $K$/atom quantifies the distribution of Wyckoff positions and thus reflects symmetry-related complexity, while $M_{LED}$ captures chemical and polyhedral environment diversity. Although a general trend of increasing $K$/atom with $M_{LED}$ is visible, the correlation is not one-to-one: some structures with low crystallographic complexity (low $K$) can still exhibit high $M_{LED}$ values due to diverse chemical coordination environments. This suggests that $M_{LED}$ provides a more gradual and chemically sensitive measure of structural diversity, whereas $K$/atom assigns similar values to many low-symmetry structures and yields a long-tailed distribution under the ≤ 20-atom constraint. **b** Distribution of $K$/atom values for the training set compared with structures generated by PIGEN conditioned on $K$/atom (target $K$/atom = 3). The generated distribution is shifted towards higher $K$/atom values relative to the training data but remains bound by the upper tail of the training set. This reflects the fact that, for the ≤ 20-atom regime, $K$/atom is concentrated in a narrow low-value region due to limited Wyckoff multiplicity, making high-complexity targets statistically rare and difficult to access for the PIGEN | $K$ model. In this context, $M_{LED}$ provides a wider range, smoother variation across structures, and can be inferred directly from atomic geometry by graph neural networks. It thus serves as a more flexible steering signal for controlled exploration beyond known structural motifs. (Fig. 3a).

**Extended Data Table 1. Performance of generative models under single- and multi-objective conditioning.** The table reports the percentage of generated structures (per 10,000-sample batch) that satisfy different validity criteria, corresponding to the sequential filtering stages illustrated in Fig. 4a-d (columns *A*-*D*), extended here to independent application of each criterion as well as to crystallographic complexity *K* per atom (column *K*). In addition to the models presented in the main text, we include other conditioning targets to probe interactions between properties, and ablation models that isolate specific contributions of the physics-informed loss (DiffCSP + PI) and classifier-free guidance (DiffCSP + CFG) relative to the baseline DiffCSP. Numbers in bold highlight best values, the second best are underscored

Column *A′* shows the popular structural validity criterion[50] (all interatomic distances > 0.5 Å), which all models pass at > 99.5%, underscoring its lack of discriminative power. Independent evaluation of each criterion reveals distinct failure modes that are otherwise masked in sequential filtering. For instance, compactness alone excludes many geometrically valid but chemically implausible structures, while $M_{LED}$ highlights when diversity-maximisation trades off with plausibility. Importantly, no single criterion suffices to guarantee chemical feasibility, a point further emphasised by the variability across conditioning strategies.

Standard error (± S.E.) reflects binomial statistics across 10000 samples (detailed in SI Fig. 5). Together with the main-text results, this table clarifies the effects of conditioning and validation choices, highlighting both strengths and limitations of different model setups in producing chemically plausible and structurally diverse candidates.

| Model | ***A′***<br>**Validity $d_{ij}$ > 0.5 Å, %** | ***A***<br>**Validity SPP < 0.36, %** | ***B***<br>**Novel composition, %** | ***C***<br>***0.55<C*** **<0.85, %** | ***D***<br>***$M_{LED}$* > 9, %** | ***K***<br>***K* > 3.9, %** |
|---|---|---|---|---|---|---|
| Pseudo-random | **100.0± 0.0** | 5.1 ± 0.4 | **90.6 ± 0.6** | 11.2 ± 0.6 | **76.6 ± 0.8** | 14.5 ± 0.7 |
| DiffCSP | 99.9 ± 0.1 | 82.8 ± 0.7 | 73.6 ± 0.9 | 61.8 ± 1.0 | 47.0 ± 1.0 | 34.0 ± 0.9 |
| MatterGen | **100.0± 0.0** | 87.9 ± 0.7 | 65.2 ± 1.0 | 67.4 ± 0.9 | 37.8 ± 1.0 | 27.7 ± 0.9 |
| DiffCSP+PI loss | 99.8 ± 0.1 | 84.6 ± 0.7 | 72.5 ± 0.9 | 64.4 ± 1.0 | 42.4 ± 1.0 | 28.3 ± 0.9 |
| **Models conditioned on target properties** | | | | | | |
| DiffCSP+CFG \| ($M_{LED}$ = 9, *C* = 0.7) | 99.7 ± 0.1 | 78.0 ± 0.8 | 74.2 ± 0.9 | 84.6 ± 0.6 | 67.2 ± 0.9 | 31.4 ± 0.9 |
| PIGEN \| $E_{hull}$ = 0 | 99.6 ± 0.1 | 85.3 ± 0.7 | 68.4 ± 0.9 | 60.9 ± 1.0 | 43.1 ± 1.0 | 30.1 ± 0.9 |
| PIGEN \| *C* = 0.7 | 99.9 ± 0.1 | 83.0 ± 0.8 | 69.2 ± 0.9 | **98.3 ± 0.3** | 41.3 ± 1.0 | 29.5 ± 0.9 |
| PIGEN \| $M_{LED}$ = 9 | 99.6 ± 0.1 | 80.0 ± 0.8 | 81.6 ± 0.8 | 67.6 ± 0.9 | 72.9 ± 0.9 | 37.3 ± 1.0 |
| PIGEN \| *K* = 3 | 99.6 ± 0.1 | 78.4 ± 0.8 | 81.1 ± 0.7 | 38.9 ± 0.9 | 52.0 ± 1.0 | 60.2 ± 0.9 |
| PIGEN \| ($M_{LED}$ = 9, $E_{hull}$ = 0) | 99.6 ± 0.1 | 83.5 ± 0.7 | 73.8 ± 0.9 | 68.9 ± 0.9 | 70.7 ± 0.9 | 36.0 ± 1.0 |
| PIGEN \| ($M_{LED}$ = 9, *C* = 0.7) | 99.8 ± 0.1 | 78.4 ± 0.8 | 75.1 ± 0.9 | 89.2 ± 0.6 | 69.7 ± 0.9 | 32.1 ± 0.9 |
| PIGEN \| (*K* = 3, *C* = 0.7) | 99.9 ± 0.1 | 77.3 ± 0.9 | 72.8 ± 0.9 | 94.1 ± 0.4 | 47.1 ± 1.0 | **61.3 ± 1.0** |
| MatterGen \| ($M_{LED}$ =9, *C*=0.7) | **100.0± 0.0** | **89.3 ± 0.7** | 59.2 ± 1.0 | 94.5 ± 0.4 | 33.0 ± 1.0 | 24.8 ± 0.9 |

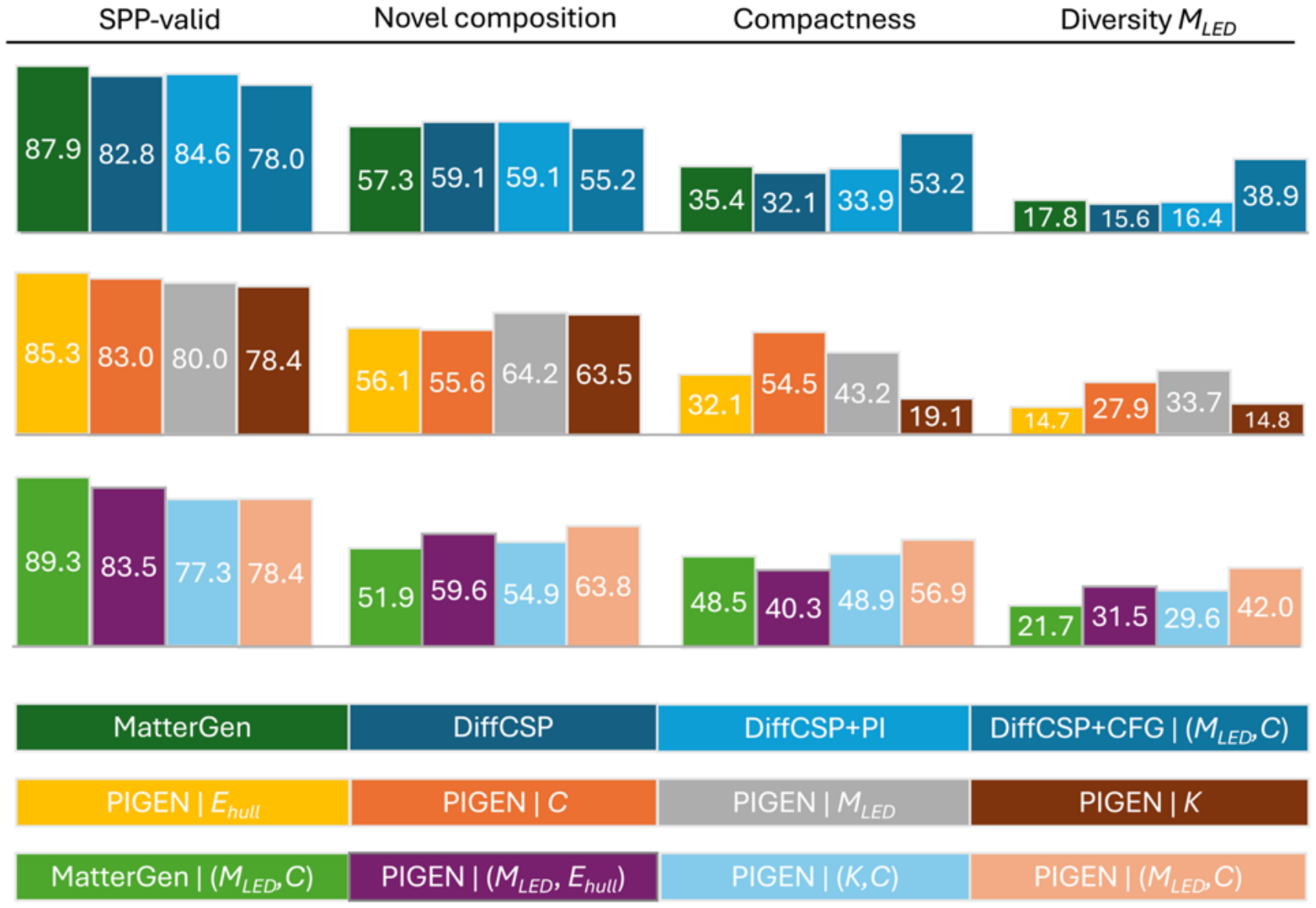

**Extended Data Figure 2. Sequential application of compositional and structural criteria for selecting plausible, novel and diverse structures.** The top row evaluates 10,000 structure batches generated with original DiffCSP, MatterGen and DiffCSP equipped with physics-informed loss (DiffCSP+PI), which demonstrate a comparable productivity, and DiffCSP+CFG conditioned on ($M_{LED}$, $C$) demonstrating an improved productivity, measured in per cent of entries with unique compositions where all elemental pairs have been experimentally confirmed, across all criteria. The "DiffCSP+PI loss" model improves DiffCSP results in all categories and outperforms MatterGen when SPP-validity and compositional novelty are both enforced. Multi-property conditioning of DiffCSP with classifier-free guidance (CFG) on $M_{LED}$ and Compactness further increases the fraction of generated structures passing all metrics. The middle row displays outcomes for PIGEN variants with single-property conditioning: $E_{hull}$, Compactness ($C$), $M_{LED}$, structural complexity $K$, illustrating how targeted conditioning shifts plausibility and diversity characteristics. The bottom row presents results for multi-property conditioned models: PIGEN | ($M_{LED}$, $E_{hull}$), PIGEN | ($K$, $C$), PIGEN | ($M_{LED}$, $C$), as well as MatterGen, fine-tuned and conditioned for ($M_{LED}$, $C$) following the adaptor approach proposed in the original study[7]. which increases the number of plausible and diverse structures. This highlights how joint conditioning, specifically on ($M_{LED}$, $C$), improves the identification of plausible, novel and diverse structures across models.

**Extended Data Table 2. Performance of a broader set of generative models** (in addition to those presented in Table 1) with application of all criteria ***A*–*E*** (Fig. 4a-e) for selecting plausible and diverse structures, energetics and maximum diversity: best performers are in bold, second best are underlined. In Column *E*, cumulative standard error after ***A*–*E*** filters ranges from 0.1–0.3%, individual SEs are omitted from the table cells for clarity.

| | ***D*** | ***E*** | ***F*** |
|---|---|---|---|
| **Model** | ***A*** & ***B*** & ***C*** & ***D*** % | ***D*** & Stable: $E_{hull}^{DFT}$ < 100 / 50 / 35 / 25 meV/atom, % | $E_{hull}^{DFT}$ < 50 meV/atom & max($M_{LED}$) |
| Pseudo-random | 0.03 | 0 / 0 / 0 / 0 | - |
| DiffCSP | 15.6 ± 0.5 | 4.0 / 1.3 / 0.7 / 0.5 | 10.7 |
| MatterGen | 16.4 ± 0.5 | 3.4 / 1.2 / 0.7 / 0.4 | 10.5 |
| DiffCSP+PI loss | 17.8 ± 0.5 | 5.1 / 2.5 / 1.6 / 1.0 | 10.7 |
| **Models conditioned on target properties** | | | |
| PIGEN \| $E_{hull}$ = 0 | 14.7 ± 0.5 | 2.0 / 1.0 / 0.7 / 0.5 | 10.5 |
| PIGEN \| $C$ = 0.7 | 27.9 ± 0.5 | 6.8 / 2.2 / 1.4 / 1.0 | 10.0 |
| PIGEN \| $M_{LED}$ = 9 | 33.7 ± 0.5 | 6.7 / 2.7 / 1.5 / 1.0 | 10.7 |
| PIGEN \| $K$ = 3 | 14.8 ± 0.5 | 1.4 / 0.7 / 0.4 / 0.2 | 10.5 |
| PIGEN \| ($M_{LED}$ = 9, $E_{hull}$ = 0) | 31.5 ± 0.5 | 1.8 / 0.8 / 0.5 / 0.4 | 10.5 |
| **PIGEN \| ($M_{LED}$ = 9, $C$ = 0.7)** | **42.0 ± 0.5** | **8.7 / 3.8 / 2.3 / 1.5** | **10.9** |
| PIGEN \| ($K$ = 3, $C$ = 0.7) | 29.6 ± 0.5 | 2.9 / 1.2 / 0.7 / 0.5 | 10.5 |
| MatterGen \| ($M_{LED}$ = 9, $C$ = 0.7) | 21.7 ± 0.5 | 6.7 / 2.7 / 1.8 / 1.2 | 10.7 |

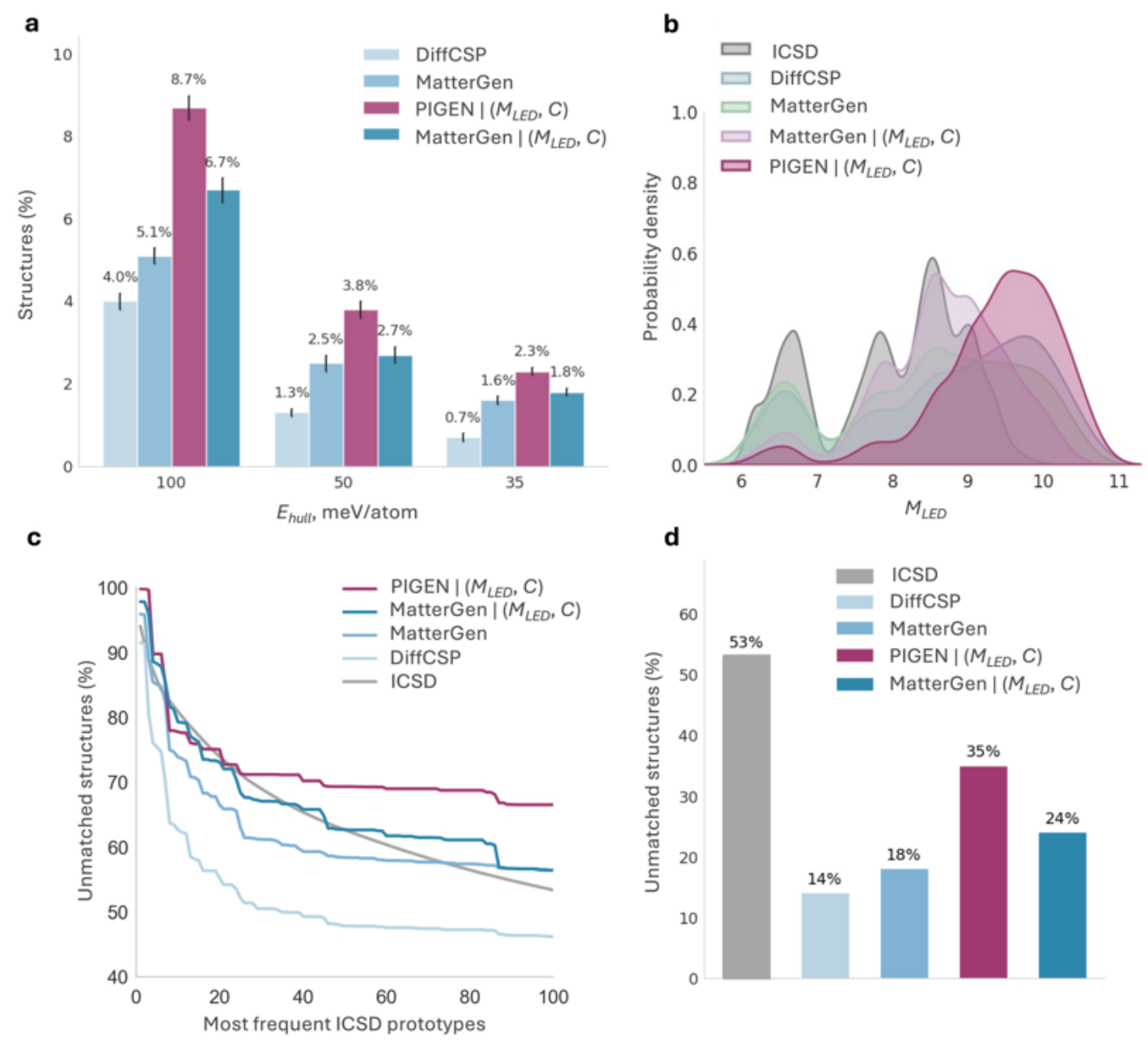

**Extended Data Figure 3. a** Fraction of generated structures from each model: DiffCSP, MatterGen, PIGEN | ($M_{LED}$, $C$), and MatterGen | ($M_{LED}$, $C$) that in addition to satisfying validation criteria ***A***–***D*** (SPP, compositional novelty, compactness, and diversity $M_{LED}$), are predicted by DFT to lie close to the convex hull with decreasing energy threshold for proximity; error bars indicate the cumulative standard error of the passing fraction after applying all filters ***A***–***E***; **b** Distributions of $M_{LED}$ for the training structures and for structures generated by PIGEN | ($M_{LED}$, $C$) and baseline generative models. DiffCSP and unconditioned MatterGen reproduce the dominant modes of the training distribution but still place 29% and 19% of samples, respectively, in the low-probability tail beyond the 99th percentile. Conditioning MatterGen on $M_{LED}$ reduces this fraction to 11%, concentrating sampling on less extreme but more populated tail regions, while increasing the fraction of valid and chemically plausible samples (panel **a**, Table 1), indicating a shift from less targeted tail sampling towards higher-quality structural diversity. In contrast, PIGEN | ($M_{LED}$, $C$) shows the strongest and most structured out-of-distribution (OOD) behaviour, with over 43% of samples lying in regions where the training distribution is lowest, corresponding to targeted exploration of rare local environments. We adopt the standard definition of OOD used in statistical novelty detection and generative modelling – namely, enrichment of samples in low-probability regions of the training distribution rather than extension beyond the empirical min-max range[36,37]. Under this criterion, PIGEN | ($M_{LED}$, $C$) achieves the strongest OOD sampling, while MatterGen | ($M_{LED}$, $C$) provides controlled, higher-quality diversity relative to default MatterGen consistent with its performance trend in **a**. **c**; Inverse cumulative match analysis of generative model outputs against common ICSD prototypes. For each generative model, the batch of generated structures was first filtered for chemical plausibility and structural diversity (criteria ***A***–***D***) and then compared to the top 100 most frequently observed structural prototypes in the ICSD. The y-axis shows the percentage of filtered structures that remain unmatched, while the x-axis lists the ICSD prototypes ranked by decreasing frequency. The grey reference curve shows the ICSD matched to itself, which decreases smoothly because all prototypes are present. In contrast, the generative model curves decrease in stepwise fashion: when no structure in a model batch matches a given prototype, the curve remains horizontal. The extent to which each curve remains above zero at $x = 100$ indicates the fraction of structures that do not match any of the top 100 ICSD prototypes, illustrating the diversity of frameworks generated beyond structural types predominant in the training data; **d** Fraction of all generated structures (per 10,000-sample batch) that, after applying the same plausibility and novelty checks (***A***–***D***), do not match any of the 100 most frequent ICSD prototypes (Methods). The top-100 prototypes account for 47% of ICSD structures (grey reference bar). Conditioning on ($M_{LED}$, $C$) (i.e. PIGEN | ($M_{LED}$, $C$) and MatterGen | ($M_{LED}$, $C$)) consistently increases the proportion of generated structures associated with less frequent structural motifs relative to hull-energy–guided sampling (i.e., MatterGen, DiffCSP). This indicates that $M_{LED}$ guidance shifts sampling into underrepresented regions of structural-environment space.

## Data Availability

All datasets used in this study are derived from publicly available crystal structure repositories (Materials Project and Alexandria) and processed following the procedures described in Methods. Preprocessed datasets, along with scripts for feature generation, are available www.github.com/lrcfmd/pigen. Crystal structures generated by the physics-informed generative models described in this work are available via the University of Liverpool Data Repository at https://doi.org/10.17638/datacat.liverpool.ac.uk/3057

## Code Availability

The source code implementing the diffusion-based generative model, evaluation metrics ($M_{LED}$), and all experiments is available at www.github.com/lrcfmd/pigen and https://doi.org/10.5281/zenodo.17357919 under an open-source license. Detailed instructions for reproducing the training and evaluation pipeline, including hyperparameter configurations and environment specifications, are provided in the repository's README and `environment.yaml`. FUSE is available at https://github.com/lrcfmd/FUSE-stable. To ensure full reproducibility, random seeds and configuration files used in the reported experiments, and model checkpoints are included.

## Acknowledgements

We acknowledge funding from the UK Engineering and Physical Sciences Research Council (EPSRC) under grant EP/V026887 and the Leverhulme Trust via the Leverhulme Research Centre for Functional Material Design (RC-2015-036). This work made use of computational resources provided by Isambard-AI, the UK's national AI supercomputer hosted by the Bristol Centre for Supercomputing at the University of Bristol, under grant BYYG-VXQ2-C.